\documentclass[twocolumn,aps]{revtex4}
\usepackage{graphicx}
\begin{document}
\draft

\title{Macroscopically ordered state in exciton system}
\author{L.~V.~Butov$^{1,2}$, A.~C.~Gossard$^3$, and D.~S.~Chemla$^{1,4}$}
\address{$^1$Materials Sciences Division, E.~O.~Lawrence
Berkeley National Laboratory, Berkeley, California 94720\\
$^2$Institute of Solid State Physics, Russian Academy of Sciences,
142432 Chernogolovka, Russia\\
$^3$Department of Electrical and Computer Engineering, University
of California, Santa Barbara, CA 93106\\
$^4$Department of Physics, University of California at Berkeley,
Berkeley, California 94720}\maketitle

\vskip1cm \noindent {\bf Macroscopically ordered arrays of
vortices in quantum liquids, such as superconductors, He--II, and
atom Bose--Einstein Condensates (BEC), demonstrate macroscopic
coherence in flowing superfluids
\cite{Trauble,Yarmchuk,Madison,Abo-Shaeer}. Despite of the rich
variety of systems where quantum liquids reveal macroscopic
ordering, experimental observation of a macroscopically ordered
electronic state in semiconductors has remained a challenging
unexplored problem. A system of excitons is a promising candidate
for the realization of macroscopic ordering in a quantum liquid in
semiconductors. An exciton is a bound pair of an electron and a
hole. At low densities, it is a Bose quasi-particle. At low
temperatures, of the order of a few Kelvins, excitons can form a
quantum liquid, i.e., a statistically degenerate Bose gas and
eventually BEC \cite{Keldysh,Lozovik,Butov98,Butov01,Butov02}.
Here, we report the experimental observation of a macroscopically
ordered state in an exciton system.}

We studied spatially resolved photoluminescence, PL, of
quasi-two-dimensional gases of indirect excitons in
GaAs/Al$_x$Ga$_{1-x}$As coupled quantum wells, QWs (Fig.~1b).
Previous studies \cite{Butov98,Butov01,Butov02} have shown that
coupled QWs is a unique system where a cold exciton gas, and, more
generally, a cold gas of light boson quasiparticles, can be
created. The indirect excitons in coupled QWs are characterized by
high cooling rates, three orders of magnitude higher than in bulk
GaAs, and long lifetime against electron--hole recombination, more
than three orders of magnitude longer than in a single GaAs QW.
This lifetime is much longer than the characteristic time scale
for cooling of initially hot photogenerated excitons down to
temperatures well below 1K, where the dilute
quasi--two-dimensional Bose gas of indirect excitons becomes
statistically degenerate \cite{Butov01}. Because the exciton mass,
$M$, is small, even smaller than the free electron mass $m_0$, the
quantum degeneracy temperature $T_0=(\pi \hbar^2 n)/(2Mgk_B)$ ($g$
is the spin degeneracy of the exciton state and $k_B$ is the
Boltzmann constant) exceeds 1 K at experimentally accessible
exciton densities, $n$, i.e. is several orders of magnitude higher
than for atoms.

\begin{figure}
    \centering
    \includegraphics[width=0.79\columnwidth]{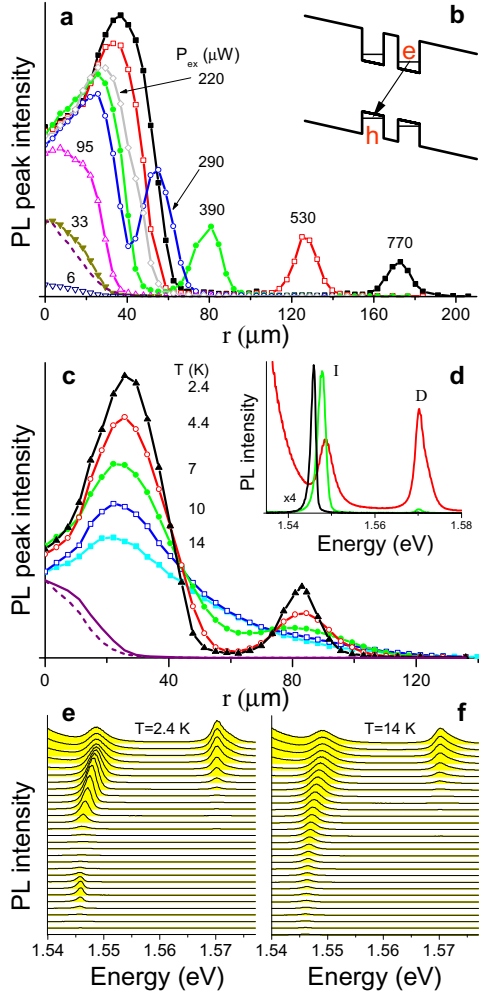}
    \caption{Radial dependence of the indirect exciton PL. {\bf a},
Peak intensity of the indirect exciton PL vs $r$, the distance
from the excitation spot center, at $T=1.8$ K, $V_g=1.22$ V, and
the excitation powers $P_{ex}=6, 33, 95, 220, 290, 390, 530,$ and
770 $\mu$W. {\bf c}, Peak intensity of the indirect exciton PL vs
$r$ at $P_{ex}=390$ $\mu$W$, V_g=1.22$ V, and $T=2.4, 4.4, 7, 10,$
and 14 K. For comparison, the solid line shows the peak intensity
variation of the direct exciton PL. The excitation spot profiles
are shown by the dotted lines in {\bf a} and {\bf c}. The
corresponding spatial dependence of the PL spectra at $T=2.4$ and
14 K are shown in {\bf e} and {\bf f}. The upper spectra
correspond to the excitation spot center, the lowest spectra are
recorded $107 \mu$m away from the excitation spot center, and the
step is $3.7 \mu$m. The indirect exciton PL line is at $\approx
1.545 - 1.55$ eV, the direct PL line is at $\approx 1.57$ eV, the
broad line arising below the indirect exciton emission comes from
the bulk $n^+ \--$GaAs emission. The selected spectra at $T=2.4$ K
are shown in ({\bf d}): the spectrum in the excitation spot center
at $r=0$ (red), the spectrum in the first ring center at $r=29
\mu$m (green), and the spectrum in the second ring center at $r=83
\mu$m (black, the spectrum intensity is multiplied by 4). {\bf b},
Energy band diagram of the CQW structures. The PL of indirect
excitons is characterized by the rings centered at the excitation
spot: the internal ring is located near the edge of the excitation
spot, while the external ring is observed far away from the
excitation spot.}
     \label{fig:1}
\end{figure}

Yet another important advantage of the system is a repulsive
interaction between the indirect excitons which, because of the
separation of the electron and hole layers (Fig.~1b), are dipoles
oriented perpendicular to the QW plane. This dipole-dipole
interaction stabilizes the exciton state against the formation of
metallic electron-hole droplets in real space
\cite{Yoshioka,Littlewood}, reinforces BEC \cite{Leggett}, and
results in a screening of an in-plane random potential (caused by
interface fluctuations, impurities, etc. and unavoidable in any
real QW sample). Note, that in ideal two-dimensional systems pure
BEC is only possible at $T=0$, although, a phase transition to a
superfluid exciton state is possible at finite temperatures
\cite{Popov}. The latter is characterized by a long-range order at
low temperatures \cite{KT}.

Typically, excitons in semiconductors are generated by a laser
photoexcitation and their density is controlled by the laser
intensity. The indirect excitons have a small oscillator strength
because of the small overlap between electron and hole
wavefunctions. Therefore, a much higher density of indirect
excitons is achieved by nonresonant laser photoexcitation with
energies at or above the direct exciton resonance where the photon
absorption coefficient is high (Fig.~1b). In a quasiequilibrium,
practically all photoexcited carriers relax to the indirect
exciton states as they are lower in energy than the direct exciton
states (the ratio between the indirect and direct exciton
densities is typically larger than $10^{4}$). The cost to pay for
the nonresonant excitation is that the initially photogenerated
excitons are hot. However, they quickly cool down to the lattice
temperature via phonon emission: e.g. the exciton temperature can
drop down to 400 mK in about 5 ns, that is the time much shorter
than the indirect exciton lifetime \cite{Butov01}. Therefore,
there are two ways to overcome the obstacle of hot generation and
study cold gases of indirect excitons with effective temperatures
close to that of the lattice: (1) use a discrimination in time and
study the indirect excitons a few ns after the end of the
photoexcitation pulses \cite{Butov01}, (2) use a discrimination in
space and study the indirect excitons beyond the photoexcitation
spot. In the latter case, excitons can cool down to the lattice
temperature as they travel away from photoexcitation spot.

Here, exploring the spatially and spectrally resolved PL
experiments, we have observed a ring structure in the indirect
exciton photoluminescence and a macroscopically ordered state of
indirect excitons appearing in the ring the most remote from the
excitation spot. First we present a brief description of our
experimental findings, that is followed by a discussion of the
observed effects.

At the lowest excitation powers, $P_{ex}$, the spatial profile of
the indirect exciton PL intensity practically follows the laser
excitation intensity (Fig.~1a). However, at high $P_{ex}$, we
observed a nontrivial pattern for that profile. First, we address
its radial dependence detailed in Fig.~1. The pattern is
characterized by a ring structure: the laser excitation spot is
surrounded by two concentric bright rings separated by an annular
dark interring region (the ratio between the indirect exciton PL
intensities in the external ring and in the dark interring region
reaches a factor of 30 at high $P_{ex}$). The rest of the sample
outside the external ring is dark. The internal ring appears near
the edge of the laser excitation spot, and the external ring can
be remote from the excitation spot by more than 100 $\mu$m. Its
radius increases with increasing excitation power. The ring
structure follows the laser excitation spot when it is moved over
the whole sample area. This nontrivial spatial profile of the
indirect exciton PL intensity is only observed at low
temperatures. When the temperature is increased the bright rings
wash out, the PL intensity in the interring region and outside the
external ring increases, and the spatial profile intensity
approaches a monotonic bell-like shape (Fig.~1c).

The azimuthal dependence of the indirect exciton PL intensity in
the external ring is also nontrivial: that ring is fragmented into
circular-shape structures that form a periodic array, Figs.~2a--e.
These fragments follow the external ring either when the
excitation spot is moved over the sample area or when the ring
radius varies with $P_{ex}$. The bright fragments always keep the
circular shape with equal dimensions in radial and azimuthal
directions under all experimental conditions studied. As already
mentioned, they form a nearly periodic chain over macroscopic
lengths, up to $\sim 1$ mm. This is demonstrated in Fig.~3e which
shows the nearly linear dependence of the fragment positions along
the ring vs their number. Along the whole external ring, both in
the peaks and the passes, the indirect exciton PL lines are
spectrally narrow with the full width at half maximum, FWHM,
$\approx 1.3$ meV, considerably smaller than in the center of the
excitation spot, Figs.~2f and 1d. The ring fragmentation is
observed at the lowest temperatures only and is already absent at
$T \approx 4$ K, Figs.~3a-c. The PL contrast along the ring washes
out as the temperature increases, this can be quantified by the
amplitude of the Fourier transform of the variation of the PL
intensity along the ring as shown in Fig.~3d.

\begin{figure}
    \centering
    \includegraphics[width=0.75\columnwidth]{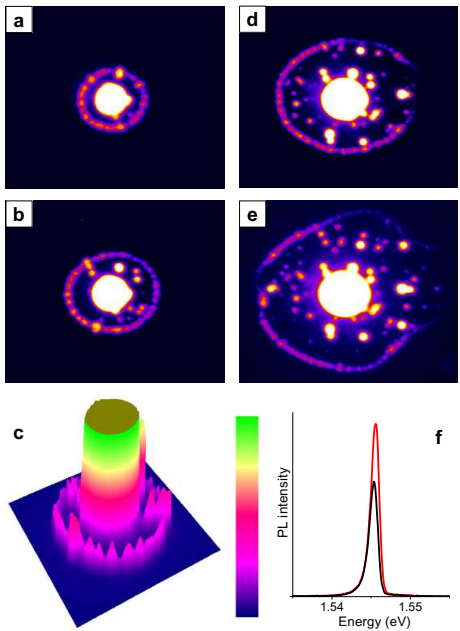}
    \caption{Spatial pattern of the indirect exciton PL intensity at
$T=1.8$ K, $V_g=1.22$ V, and $P_{ex}=290 ({\bf a}), 390 ({\bf b,
c}), 690 ({\bf d}),$ and 1030 ({\bf e}) $\mu$W. On {\bf a, b, d,
e}, the area of view is $530 \times 440 \mu$m. The external ring
of the indirect exciton PL is fragmented into a periodic chain of
circular structures. The fragments follow the external ring both
when its radius is changed by varying the excitation intensity or
when the laser excitation is moved on the sample. Besides these
mobile features, the indirect exciton PL intensity is also
strongly enhanced in the certain spots within the area terminated
by the external ring. The position of these spots is fixed on the
sample. {\bf f}, Indirect exciton PL spectra in a peak (red) and
the adjacent pass (black) on the fragment chain along the ring.}
     \label{fig:2}
\end{figure}

Besides the mobile features, like the rings and the external
ring's fragments that move with the excitation spot or when
$P_{ex}$ is varied, the spatial pattern shows also that the
indirect exciton PL intensity is strongly enhanced in certain
fixed spots on the sample, Figs.~2a--e. We call them localized
bright spots (LBS). For any excitation spot location and any
$P_{ex}$ the LBS are only observed when they are within the area
terminated by the external ring, Figs.~2a--e. In the LBS the
indirect exciton PL line is spectrally narrow, FWHM $\approx 1.2$
meV, and its energy is locally reduced. The LBS also wash out with
increasing temperature.

\begin{figure}
    \centering
    \includegraphics[width=0.7\columnwidth]{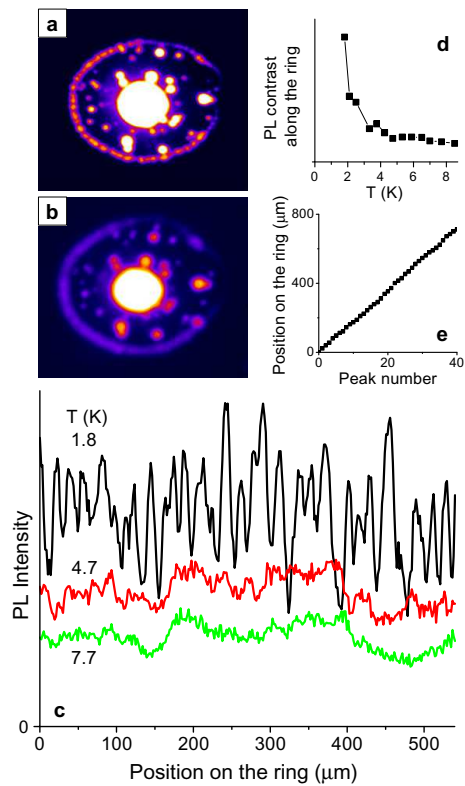}
    \caption{Temperature dependence of the spatial pattern of the
indirect exciton PL intensity. {\bf a, b}, The pattern at $T=1.8$
({\bf a}) and 4.7 K ({\bf b}) for $V_g=1.22$ V, and $P_{ex}=690
\mu$W. The area of view is $475 \times 414 \mu$m. {\bf c}, The
corresponding variation of the indirect exciton PL intensity along
the external ring at $T=1.8, 4.7,$ and 7.7 K. The ring
fragmentation into the periodic chain washes out with increasing
temperature. This is visualized by the PL contrast ({\bf d})
presented by an amplitude of the Fourier transform. The dependence
of the position of the indirect exciton PL intensity peaks along
the external ring vs the peak number is nearly linear ({\bf e}),
showing that the fragments form a periodic chain.}
     \label{fig:3}
\end{figure}

All these effects are observed on several mesas studied and all
experimental data are reproducible after cycling the sample
temperature up to room temperature and back to 1.8 K many times.

We now discuss the nature of the observed effects. Under cw laser
excitation a system of photoexcited excitons freely expanding in a
QW plane is a thermodynamically open system in a quasiequilibrium.
At low densities the indirect excitons are localized by the
in-plane potential fluctuations and, therefore, the spatial
profile of the indirect exciton PL intensity practically follows
the laser excitation intensity, Fig.~1a. At high densities,
however, the repulsive interaction between the indirect excitons
screens the in-plane potential fluctuations and, therefore,
results in a delocalization of the indirect excitons. Their long
lifetime allows them to move far away from the excitation spot
before they recombine. As we show below, the lifetime is further
enhanced by the exciton motion itself. All these factors
facilitate the exciton transport over macroscopic distances, up to
$\sim 1$ mm, in our experiments.

It is well known that for delocalized quasi-two-dimensional
excitons only the states inside the radiative zone terminated by
the photon cone, Fig.~4a, can recombine radiatively by resonant
emission of photons \cite{Feldmann}. Those are the states with
small in-plane center of mass momenta $K_{\parallel} \le
K_0=E_g\sqrt{\varepsilon}/(\hbar c)$ ($E_g$ is the band gap and
$\varepsilon$ is the dielectric constant). The exciton radiative
decay rate and, therefore, the exciton PL intensity are determined
by the fraction of excitons inside the radiative zone. In the
center of the excitation spot the exciton gas is characterized by
a high effective temperature, larger than that of the lattice
\cite{Butov01}. Under cw photoexcitation, there is a continuous
flow of excitons out of the excitation spot due to the exciton
drift and diffusion (other mechanisms such as ballistic transport
and phonon wind may also be contributing to the exciton cloud
expansion). The exciton diffusion originates directly from the
exciton density gradient. The exciton drift also originates from
the density gradient as the latter gives rise to the gradient of
the indirect exciton energy because of the repulsive interaction
(see Figs.~1e, f). As the excitons travel away from the excitation
spot, the effective exciton temperature, $T_X$, decreases with
increasing the radial distance due to their energy relaxation. The
reduction of $T_X$ increases the radiative zone occupation and,
therefore, increases the PL intensity, that is seen as onset of
the internal ring. The internal ring is therefore the spatial
analog of the PL intensity enhancement after an excitation pulse,
i.e. the PL jump, observed in Ref.~\cite{Butov01}. Estimates for
the exciton density in the internal ring exceed $3 \times 10^{10}$
cm$^{-2}$ at the highest $P_{ex}$, implying that a statistically
degenerate Bose gas of indirect excitons forms in the internal
ring (at $n=3 \times 10^{10}$ cm$^{-2}$ and $T=2$ K, the
Bose-Einstein distribution function gives the occupation number of
the lowest energy state $\nu=e^{T_0/T}-1 \approx 0.3$ for the
excitons in our coupled QWs where $g=4$ and $M=0.21 m_0$).

\begin{figure}
    \centering
    \includegraphics[width=1.17\columnwidth]{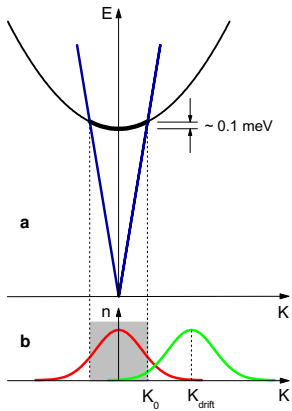}
    \caption{Schematics demonstrating a reduction of emission
intensity for excitons in motion. {\bf a}, Energy diagram for the
exciton and photon dispersion, the bold sector of the exciton
dispersion indicates the radiative zone. {\bf b}, The schematic
momentum distribution of excitons without (red) and with (green)
average drift velocity. The exciton radiative decay rate is
proportional to the fraction of excitons in the radiative zone
(grey area).}
     \label{fig:4}
\end{figure}

To explain the dark region between the rings we propose the
following scenario (Fig.~4). Travelling out of the energy ``hill''
at the center of the excitation spot excitons acquire an average
drift momentum. As the height of the energy hill, several meV
(Fig.~1e), is much larger than the radiative zone energy width,
$\sim 0.1$ meV, the average momentum acquired, $K_{drift}$, can
exceed $K_0$. That means that the moving excitons move out of the
radiative zone and, therefore, become optically inactive. This
explains the existence of the dark region between the internal and
external rings. An interesting aspect of this effect is that the
optically inactive excitons move at speeds several times faster
than the speed of sound: already for $K=K_0$ excitons their speed
$v=\hbar K_0 /M=1.4 \times 10^6$ cm/s is larger than the speed of
sound in GaAs $v_s=3.7 \times 10^5$ cm/s. This implies that the
exciton flow in the dark interring region is diffusive rather than
superfluid, because the Landau criterion for superfluidity is not
fulfilled for excitons moving with supersonic velocities (the
exciton moving with $v \ge v_s$ can scatter to a lower energy
state with an acoustic phonon emission, and this process causes a
dissipation). Far from the excitation spot the main driving force
for exciton transport, the energy gradient, vanishes and they
relax down to the lowest energy states. This results in the sharp
enhancement of the radiative zone occupation and, therefore, the
PL intensity, that is seen as the external ring. As excitons in
the external ring relax down to the low momentum states,
characterized by low velocities, the exciton flow is stopped and
there are practically no excitons outside of that ring
(Figs.~2a--e).

The most interesting feature of the external ring is its
fragmentation into a periodic array. The existence of this
periodic ordering shows that the exciton state formed in the
external ring has a coherence on a macroscopic length scale. We
emphasize that the coherence is not driven by a laser excitation
because in our experiment the photoexcited carriers experience
multiple inelastic scatterings before the optically active
indirect excitons are formed. Instead, the coherence spontaneously
appears in the exciton system. The understanding of the
microscopic nature of this ordered exciton state warrants future
studies. We suggest that the fragments are vortices in the exciton
system and that the ordering is the consequence of a repulsive
interaction between the vortices. This is confirmed by the fact
that the fragments are always of a circular shape. The vortex
rotation is continuously supported by the exciton flow out of the
excitation spot. Similar to arrays of vortices in atom BEC
\cite{Abo-Shaeer}, the rotation can be initiated even without an
apparent external torque. A possible cause for the initiation of
the rotation is a small deviation from axial symmetry for the
exciton flow due to the in-plane potential fluctuations, which
could result in a branching of the exciton flow, e.g. similar to
the branching observed for the electron flow \cite{Topinka}. We
note also that a spontaneous macroscopic flow organization with
periodic vortical structures is a general property of
thermodynamically open systems described by nonlinear partial
differential equations
\cite{Trauble,Yarmchuk,Madison,Abo-Shaeer,Taylor}.

In-plane potential fluctuations could influence also position of
the vortices on the external ring due to the pinning effect.
However, the pinning effect appears to be small so that vortices
remain free to move (e.g. with changing of the excitation spot
location on the sample) and the deviation of the vortex position
out of the periodic array due to the pinning force is small
(Fig.~3e). On the contrary, potential fluctuations can play a
crucial role in formation of the LBS observed inside the external
ring. Based on the current data, we suggest that in the LBS the
optically inactive moving indirect excitons are captured by a
potential trap formed by in-plane potential fluctuations, relax to
the optically active exciton states, and recombine giving rise to
the spectrally narrow emission seen as the LBS. The LBS will be
studied in details later. In the context of the model discussed
here, we note: (i) the existence of LBS in the dark annular region
between the two rings confirms the presence of the optically
inactive excitons in this region and (ii) the absence of LBS
outside of the external ring confirms the absence of excitons
there.

Note that the spatial profiles of the direct exciton PL intensity
exhibit none of the above effects and practically follows the
laser excitation profile for all temperatures and excitation
densities studied, Fig.~1c (this is also the case for the bulk
GaAs emission). This is consistent with a short lifetime of direct
excitons that limits the distance they can travel before the
recombination and does not allow an effective cooling for them.\\

\noindent {\bf Methods}

\noindent {\small $n^+$-$i$-$n^+$ GaAs/AlGaAs CQW structure was
grown by MBE. The $i$-region consists of two 8nm GaAs QWs
separated by a 4nm Al$_{0.33}$Ga$_{0.67}$As barrier and surrounded
by two 200nm Al$_{0.33}$Ga$_{0.67}$As barrier layers. The electric
field in the sample growth direction is monitored by the external
gate voltage $V_g$ applied between the highly conducting
$n^+$-layers. The narrow PL linewidth indicates a small in-plane
disorder. For $V_g \approx 0$ the ground state of the optically
pumped CQW is the direct exciton made of electron and hole in the
same layer and similar to the excitons in single QWs. For nonzero
$V_g$ the ground state is the indirect exciton made of electrons
and holes in different layers, Fig.~1b. The indirect exciton
lifetime is in the range of tens and hundreds of ns. The indirect
exciton energy shift with density, $\delta E(n)$, allows us to
evaluate their concentration using $\delta E(n) = 4 \pi n e^2
d/\varepsilon$, where $d$ is the effective separation between the
electron and hole layers. Radiative recombination is the dominant
decay mechanism of indirect excitons in our high-quality sample.
The sample was excited with a HeNe laser, $\lambda=632.8$nm. The
experiments were performed in He$^4$ cryostat with optical
windows. Spatially resolved PL spectra were detected using a
pinhole in the intermediate image plane. In the image experiment,
the spatial pattern of the indirect exciton PL intensity was
detected by CCD camera using spectral selection of the indirect
exciton emission by the interference filter adjusted to the
indirect exciton energy. The spatial resolution was $5 \mu$m.}

\noindent {\small {\bf Acknowledgements}\\
We thank A.L. Ivanov for discussions, A.V. Mintsev and C.W. Lai
for help in preparing the experiment, K.L. Campman for growing the
high quality coupled QW samples. This work was supported by the
Director, Office of Science, Office of Basic Energy Science,
Division of Materials Sciences, U.~S.~Department of Energy under
Contract No.~DE-AC03-76SF00098 and by the RFBR.}

\end{document}